\documentclass[useAMS,usenatbib,onecolumn]{mnras}

\usepackage{dcolumn}
\usepackage{physics}
\usepackage{adjustbox}
\usepackage{graphicx}
\usepackage{verbatim}
\usepackage{tikz}
\usepackage{caption}
\usepackage[version=4]{mhchem}
\usepackage{xcolor}
\usepackage{url}
\usepackage{color}

\usepackage{amssymb}

\usepackage{epsf}
\usepackage{amsmath}
\usepackage{lscape}	
\usepackage{bm}

\usepackage{siunitx}

\usepackage{booktabs}
\usepackage{multirow}

\usepackage{dcolumn}
\newcolumntype{d}[1]{D{.}{.}{#1}}

\newcommand{\cm}{cm$^{-1}$}

\newcommand{\ai}{\textit{ab initio}}

\newcommand{\TROVE}{{\sc TROVE}}
\newcommand{\Marvel}{{\sc Marvel}}
\newcommand{\MARVEL}{{\sc Marvel}}

\newcommand{\name}{Dozen}

\graphicspath{{pdf/}}

\newcommand{\COO}[2]{$^{1#1}$C$^{1#2}$O$_2$}
\newcommand{\OCO}[3]{$^{1#1}$O$^{1#2}$C$^{1#3}$O}

\newcommand{\2}{$_2$}

\title[ExoMol line lists -- LXIII: Isotopologues of CO$_2$]{ExoMol line lists -- LXIII: ExoMol line lists for 12 isotopologues of CO$_2$}

\date{\today}

\author[]{ 
Sergei N. Yurchenko\thanks{The corresponding author: s.yurchenko@ucl.ac.uk},
Marco G. Barnfield,
Charles A. Bowesman,
Ryan P. Brady,
\newauthor
Elizabeth R. Guest,
Kyriaki Kefala,
Qing-He Ni,
Armando N. Perri,
Oleksiy A. Smola,
\newauthor
Andrei Solokov, Chenyi Tao,
Jonathan Tennyson\thanks{The corresponding author: j.tennyson@ucl.ac.uk},\\
Department of Physics and Astronomy, University College London, Gower Street, WC1E 6BT London, United Kingdom
}

\date{Accepted XXXX. Received XXXX; in original form XXXX}

\pagerange{\pageref{firstpage}--\pageref{lastpage}} \pubyear{2024}

\begin{document}

\label{firstpage}

\maketitle

\begin{abstract}

Extensive rovibrational line lists are constructed for 12 isotopologues of carbon dioxide: \COO{2}{6}, \COO{3}{6}, \COO{2}{7}, \COO{3}{7}, \COO{2}{8}, \COO{3}{8}, \OCO{6}{2}{7}, \OCO{6}{2}{8}, \OCO{6}{3}{7}, \OCO{6}{3}{8}, \OCO{7}{2}{8}, and \OCO{7}{3}{8}. The variational program \textsc{TROVE} was employed together with an exact kinetic energy operator, accurate empirical potential energy surface (Ames-2) and the \textit{ab initio} dipole moment surface Ames-2021-40K. Empirical energy levels from the most recent MARVEL analyses, as well as from the HITRAN and CDSD databases, are used to replace calculated values where available. The line lists are further supplemented by assigning AFGL quantum numbers using machine-learning based estimators. The resulting data were employed to generate opacities with four radiative transfer codes, TauREx, ARCiS, NEMESIS, and petitRADTRANS, both for individual isotopologues and for  CO$_2$ at terrestrial isotopic natural abundance. All line lists and associated data are available at \url{www.exomol.com}.

\end{abstract}

\begin{keywords}
molecular data - opacity - planets and satellites: atmospheres - stars: atmospheres - ISM: molecules.
\end{keywords}

\newpage

\section{Introduction}

Carbon dioxide (CO$_2$) is a key atmospheric constituent across a wide range of planetary environments, including Earth \citep{17OyPaDr.CO2}, Venus \citep{15FeBeBe.CO2,14SnStGr.CO2}, and Mars \citep{24ReSi.CO2,13WeMaFl.CO2}, as well as exoplanets \citep{13OpBaBe.CO2,09SwTiVa.CO2,09SwVaTi.CO2,16HeLyxx.CO2} and brown dwarfs \citep{14FrLuFo.CO2,17LiMaLi.CO2}. In the Earth's atmosphere, CO$_2$ is the second most abundant greenhouse gas after water vapour \citep{17OyPaDr.CO2}, and is therefore central to studies in environmental chemistry. Several dedicated missions \citep{11BuGuHa.CO2,04CrAtBr.CO2,10AbRiAl.CO2} monitor Earth's atmospheric CO$_2$ content. Since CO$_2$ is a major component of the atmospheres of Earth, Mars, and Venus, remote sensing of this molecule is expected to be equally important for characterising exoplanets with similar properties. For example, CO$_2$ dominates the opacity of the Venusian atmosphere, comprising approximately 95\%\ of its composition \citep{93PoDaGr.CO2}.
The critical role of CO\2\ in atmospheric retrievals has been highlighted in multiple planetary science studies ranging from terrestrial remote sensing (e.g. \citet{21RoChPr.CO2}) to planetary atmospheres such as Mars (e.g., \citet{21JiLoFu.planets}).

In the context of exoplanetary science, CO$_2$ is anticipated to be a key molecular absorber across a broad range of planetary types, from temperate terrestrial planets to hot Jupiters. Its strong 4.3~$\mu$m band has been unambiguously detected with JWST in the atmospheres of WASP-39~b \citep{23AhAlBa.exo}, WASP-166~b \citep{25MaFoLo.exo}, and the sub-Neptune K2-18~b \citep{23MaSaCo.exo,24HoMa.exo}. Consequently, comprehensive and accurate opacity data for CO$_2$ and its isotopologues, valid over wide spectral and temperature ranges, are essential for probing atmospheric composition and physical conditions such as temperature and pressure. Such data underpin high-accuracy atmospheric models and are crucial for interpreting spectroscopic observations from ground-based telescopes and space missions, including JWST and the future ESA  space observatory Ariel \citep{jt946}.

Detection of CO$_2$ with high-resolution spectroscopy has not yet been achieved, mainly due to challenges associated with telluric contamination \citep{22CaGiGu.exo}, although searches are ongoing \citep{24BiHaRa.exo}. Isotopic ratios, in particular \COO{3}{6}/\COO{2}{6}, have been proposed as tracers of carbon isotope fractionation in exoplanetary atmospheres, with potential links to biological processes detectable by JWST \citep{23GlSePe.exo}. Recently, four isotopologues of CO$_2$  -- \COO{2}{6}, \COO{3}{6}, \OCO{6}{2}{8}, and \OCO{6}{2}{7} -- were detected in a terrestrial planet-forming region of an externally irradiated Herbig disk using VLT \citep{25FrBiRa.exo}.

Significant progress in developing high-resolution, variationally computed CO$_2$ line lists has been made in recent years by the NASA Ames group \citep{12HuScTa.CO2,14HuGaFr.CO2,17HuScFr.CO2,13HuFrTa.CO2} and the ExoMol project \citep{15ZaTePo.CO2,17ZaTePoa.CO2,17ZaTePob.CO2,jt804}. High-temperature line lists are particularly important for modelling exoplanetary atmospheres, and numerous works have targeted this goal \citep{14HuGaFr.CO2,17HuScFr.CO2,86WaRoxx.CO2,03TaPeTe.CO2,11TaPexx.CO2,jt480,13HuFrTa.CO2,jt804}. Recent high-temperature datasets include UCL-4000 \citep{jt804}, Ames-2021 \citep{22HuScFr.CO2}, the latest Ames line lists AI-3000K \citep{23HuFrTa.CO2} and its  HITEMP compilation   \citep{25HaGoHu.CO2}. An accurate room-temperature line list CDSD-2024-PI was recently reported by \citet{25KoPe.CO2}. While ExoMol and Ames line lists combine experimental results with variational calculations, improved with empirically-derived energies, the CDSD-2024-PI \citep{15TaPeGa.CO2} and HITRAN \citep{jt836,jt841} line lists rely on effective Hamiltonian models. All provide accurate and comprehensive rovibrational data for CO$_2$, but further improvements in accuracy and completeness remain necessary.

An additional aspect relevant to this work is the treatment of pressure broadening. A large body of literature exists on CO$_2$ broadening coefficients with different perturbing species; here we only highlight a few recent developments. \citet{13GaLaxx.broad} presented a detailed analysis of a large set of experimental broadening parameters for the main isotopologue colliding with N$_2$, O$_2$, and CO$_2$, with further analyses and reassessments reported by \citet{14GaLaBl.broad} and \citet{20HaGoTr.broad}.

In this work we present extensive rovibrational line lists for 12 isotopologues of CO$_2$: \COO{2}{6}, \COO{3}{6}, \COO{2}{7}, \COO{3}{7}, \COO{2}{8}, \COO{3}{8}, \OCO{6}{2}{7}, \OCO{6}{2}{8}, \OCO{6}{3}{7}, \OCO{6}{3}{8}, \OCO{7}{2}{8}, and \OCO{7}{3}{8}. The line lists were generated with \TROVE\ \citep{TROVE_prog} using an exact kinetic energy operator (KEO), the empirical potential energy surface (PES) Ames-2 \citep{17HuScFr.CO2} and the \ai\ dipole moment surface (DMS) Ames-2021-40K \citep{23HuFrTa.CO2}. To improve line position accuracy, (i) we employed the empirical band-centre correction approach \citep{jt503,jt729}, shifting the diagonal vibrational Hamiltonian terms to match experimental values, followed by (ii) substitution of calculated energy levels with empirical ones from \MARVEL\ studies \citep{jt925,jt932,jt955,jt963,jt973,jt974,jt983,jt984}, HITRAN \citep{jt836,jt841}, and CDSD-2024-PI \citep{25KoPe.CO2}. As part of this work, empirical energies of the minor CO$_2$ isotopologues were extracted from HITRAN using the \MARVEL\ procedure.

The resulting line lists, in ExoMol format \citep{jt548,jt939}, contain energies, Einstein A coefficients, lifetimes,  uncertainties, and quantum numbers for 12 isotopologues of CO$_2$. Partition functions and broadening parameters are also provided. Associated opacities were computed for the four radiative transfer codes ARCiS~\citep{ARCiS}, TauREx~\citep{TauRex3}, NEMESIS~\citep{NEMESIS} and petitRADTRANS~\citep{19MoWaBo.petitRADTRANS} using the  ExoMolOP procedure~\citep{jt801}, both for individual isotopologues and for a composite opacity assuming terrestrial isotopic abundances. All data are available from the ExoMol database at \url{www.exomol.com}.

\section{Rovibrational states of CO$_2$ isotopologues: quantum numbers and selection rules}
\label{sec:QN}

Accurate description of the rovibrational states is essential for the construction and assignment of spectroscopic line lists. Quantum numbers provide the link between the variationally computed eigenstates of CO$_2$ and the spectroscopic notation used in experimental and atmospheric studies. For CO$_2$ and its isotopologues, several labelling schemes exist, reflecting both the linear triatomic structure of the molecule and the strong Fermi-resonance interactions that complicate vibrational assignments. In addition, nuclear-spin statistics and molecular symmetry play a central role in determining the allowed rovibrational levels and transition selection rules. In the following subsections we outline the quantum number conventions, symmetry classifications, and selection rules adopted in this work, following both spectroscopic practice and the requirements of large-scale automatic assignments.

Vibrational states of symmetric (non-linear) XY$_2$ molecules can be described by three normal modes, conventionally denoted as $\nu_{1}$ (symmetric stretch), $\nu_{2}$ (bend), and $\nu_{3}$ (asymmetric stretch). For linear symmetric triatomic molecules such as CO$_2$, Herzberg's notation is commonly used to label vibrational states. The three modes are associated with vibrational quantum numbers $v_1$, $v_2^{\ell_2}$, and $v_3$. The superscript $\ell_2$ denotes the vibrational angular momentum associated with the doubly-degenerate bending mode in the linear configuration. This quantum number represents the projection of the vibrational angular momentum along the molecular axis and takes values from $-v_2$ to $+v_2$ in steps of two. In spectroscopic notation, however, the absolute value $|\ell_2|$ is often used.

The most widely adopted scheme for labelling vibrational states of CO$_2$ isotopologues is that of the Air Force Geophysics Laboratory (AFGL) \citep{65AmPi.CO2,81RoYoxx.CO2,08ToBrMi.CO2}. The AFGL notation employs a quintuplet of quantum numbers  $(m_{1},m_{2},l,m_{3},r)$ designed to simplify the description of strong Fermi resonances in CO$_2$ of states $\nu_1$ and $2\nu_2$, including their overtones and combination states, i.e. between states of the type $(v_{1},(v_{2}+2)^{\ell_2},v_{3})$ and $(v_{1}+1,v_{2}^{\ell_2},v_{3})$.
The vibrational quantum numbers ($m_1$, $m_2$, $m_3$) are identical for all members of the same resonance polyad, while the fifth index $r$ ranks the states within it. In AFGL notation, $m_2$ and $l$ are always equal, meaning $l \geq l_2$. By convention, for a given Fermi multiplet,  $r$  starts from 1 and increases by removing one quantum of $\nu_1$ and adding two quanta of $\nu_2$.
For example, the Fermi-resonant states $(2,0^{0},1)$, $(1,2^{0},1)$, and $(0,4^{0},1)$ correspond to $(2\,0\,0\,1\,1)$, $(2\,0\,0\,1\,2)$, and $(2\,0\,0\,1\,3)$, respectively \citep{08ToBrMi.CO2}. The polyad number $P = 2v_{1} + v_{2} + 3v_{3}$ is approximately conserved within a Fermi multiplet.

For asymmetric isotopologues such as \OCO{6}{2}{7}, the AFGL quantum numbers follow the same scheme, $(J, m_{1}, m_{2},  m_{3}, r, e/f)$, even though the concepts of symmetric and asymmetric stretching lose their strict meaning. While AFGL notation provides a useful framework for analysing resonance clusters, it is not well suited for the automatic assignment of very large datasets.

\subsection{Rotational quantum number and parity}

The total angular momentum of CO$_2$ is denoted by $J$, which includes contributions from both overall molecular rotation and vibrational angular momentum ($\ell_2 \neq 0$).

The Pauli principle imposes restrictions on symmetric isotopologues containing zero-spin oxygen nuclei, namely C$^{16}$O$_2$ and C$^{18}$O$_2$. For these species, vibrational states with even $v_3$ (symmetric under oxygen exchange) are only associated with even $J$, while states with odd $v_3$ (antisymmetric) only allow odd $J$. This reflects the requirement that the total rovibrational wavefunction be symmetric under exchange of the two oxygen nuclei. The rotationless parity of a state is labelled $e$ or $f$ and is determined by the sum $J + \ell_{2} + v_{3}$: states with even values correspond to $e$ parity, and those with odd values to $f$ parity.

For C$^{17}$O$_2$ (\COO{2}{7} and \COO{3}{8}), the $^{17}$O nuclei have non-zero spin ($I=5/2$). In this case the Pauli principle does not restrict odd or even levels, and both ortho and para nuclear-spin symmetries exist. All $J$ values are allowed, and bands with $\ell_{2} > 0$ exhibit $\ell$-type doubling, giving rise to both $e$ and $f$ parity levels for each $J$.

\subsection{Symmetry group classification}

An alternative classification of rovibrational states uses molecular symmetry groups \citep{98BuJe.method}, specifically C$_{2v}$(M) or C$_s$(M), and their irreducible representations (irreps). For symmetric CO$_2$, states transform according to the C$_{2v}$(M) group, which has four irreps: $A_1$, $A_2$, $B_1$, and $B_2$.

For isotopologues with zero-spin oxygen nuclei, the Pauli principle forbids rovibrational states of $B_1$ and $B_2$ symmetry, leaving only $A_1$ and $A_2$, which correlate with $e/f$ parity according to:
\begin{align}
&e: (-1)^J =  1, \quad A_1, \\
&e: (-1)^J = -1, \quad A_2, \\
&f: (-1)^{J+1} =  1, \quad A_2, \\
&f: (-1)^{J+1} = -1, \quad A_1.
\end{align}

For symmetric CO$_2$ with non-zero oxygen nuclear spin, all four irreps ($A_1$, $A_2$, $B_1$, $B_2$) are allowed. These can be correlated with ortho and para spin symmetries as summarised in Table~\ref{t:gns}. For the $e/f$ labels, the convention $p = (-1)^J$ for $e$ and $p = (-1)^{J+1}$ for $f$ is applied, where $p$ denotes molecular parity.

For asymmetric isotopologues, the rovibrational states transform as $A'$ or $A''$ in the C$_s$(M) group. Their relation to $e/f$ parity follows the same rules as in the symmetric, zero-spin case:
\begin{align}
&e: (-1)^J =  1, \quad A', \\
&e: (-1)^J = -1, \quad A'', \\
&f: (-1)^{J+1} =  1, \quad A'', \\
&f: (-1)^{J+1} = -1, \quad A'.
\end{align}

\subsection{Selection rules}

Rotational selection rules for linear molecules are $\Delta J = 0, \pm 1$. For CO$_2$, Q-branch transitions ($\Delta J=0$) occur only in parallel bands and involve $e \leftrightarrow f$ transitions, where the dipole moment change is parallel to the molecular axis. P- and R-branch transitions ($\Delta J = \pm 1$) occur in both parallel and perpendicular bands and involve $e \leftrightarrow e$ or $f \leftrightarrow f$.

For symmetric isotopologues with non-zero oxygen nuclear spin, additional nuclear-spin selection rules apply:
\[
\text{ortho} \leftrightarrow \text{ortho}, \qquad \text{para} \leftrightarrow \text{para}.
\]

In terms of molecular symmetry, these correspond to
\[
A_1 \leftrightarrow A_2, \qquad B_1 \leftrightarrow B_2.
\]

For asymmetric isotopologues, the $e/f$ selection rule is general for all dipole allowed transitions and therefore still holds; in group-theoretical notation it is expressed as
\[
A' \leftrightarrow A''.
\]

The nuclear-spin degeneracy factors for the isotopologues considered in this work are summarised in Table~\ref{t:gns}.

\begin{table}
\centering
\caption{\label{t:gns} Nuclear spin statistical weights $g_{\rm ns}$ for CO$_2$ isotopologues.
Values are given for the relevant irreducible representations of the molecular symmetry groups. These factors are applied in partition function calculations and line intensity normalisation.  }
\begin{tabular}{ccrrrrrrr}
\hline
\hline
          &          &     &    \multicolumn{6}{c}{$g_{\rm ns}$}  \\
\hline
 $ \Gamma$& p/o      &  p  &   626    &   636    &      828 &   838    &   727   &   737    \\
 \hline
 $ A_1   $& para     &  1  &        1 &        2 &        1 &        2 &      15 &      30  \\
 $ A_2   $& para     & -1  &        1 &        2 &        1 &        2 &      15 &      30  \\
 $ B_1   $& ortho    & -1  &        0 &        0 &        0 &        0 &      21 &      42  \\
 $ B_2   $& ortho    &  1  &        0 &        0 &        0 &        0 &      21 &      42  \\
 \hline
 $ \Gamma$&          &     &  627    &   637    &      628 &   638    &   728   &   738    \\
 \hline
 $ A'    $&          &  1  &        6 &       12 &        1 &        2 &       6 &      12  \\
 $ A''   $&          & -1  &        6 &       12 &        1 &        2 &       6 &      12  \\
 \hline
 \hline
\end{tabular}
\end{table}

\section{Line list production}

The generation of high-quality line lists requires three key ingredients:
(i) accurate PES and DMS to describe molecular structure and transition intensities;
(ii) robust nuclear-motion calculations to solve the rovibrational Schr\"{o}dinger equation;
and (iii) empirical refinement procedures to bring calculated energy levels into agreement with experimental data.
In this section we outline the methodology adopted for CO$_2$ isotopologues, including the choice of potential and dipole moment surfaces, the details of the variational calculations, and the empirical corrections applied to improve the accuracy of the final line lists.

\subsection{Potential Energy Surface}

\begin{figure}
\centering
\includegraphics[width=0.45\textwidth]{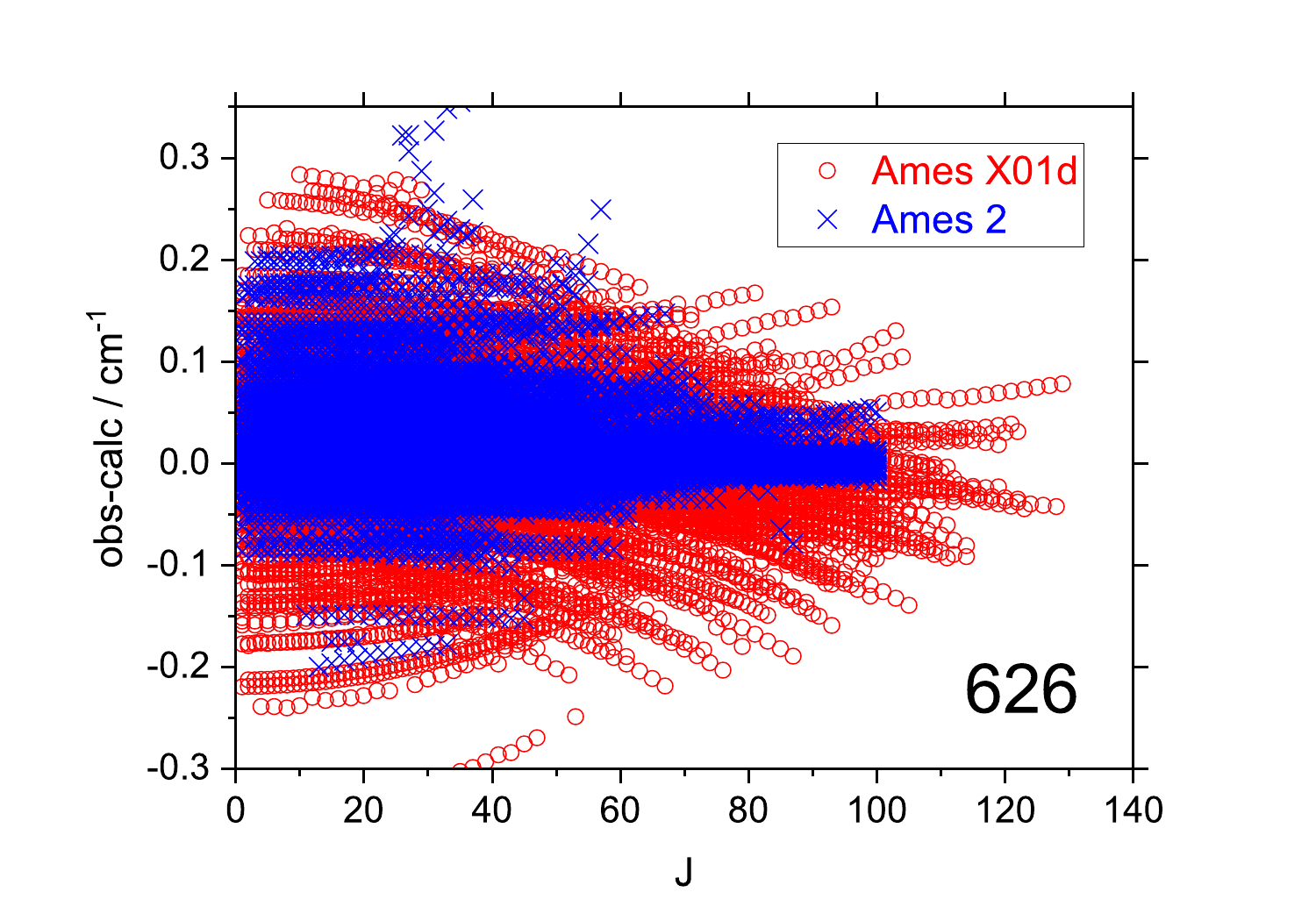}
\includegraphics[width=0.45\textwidth]{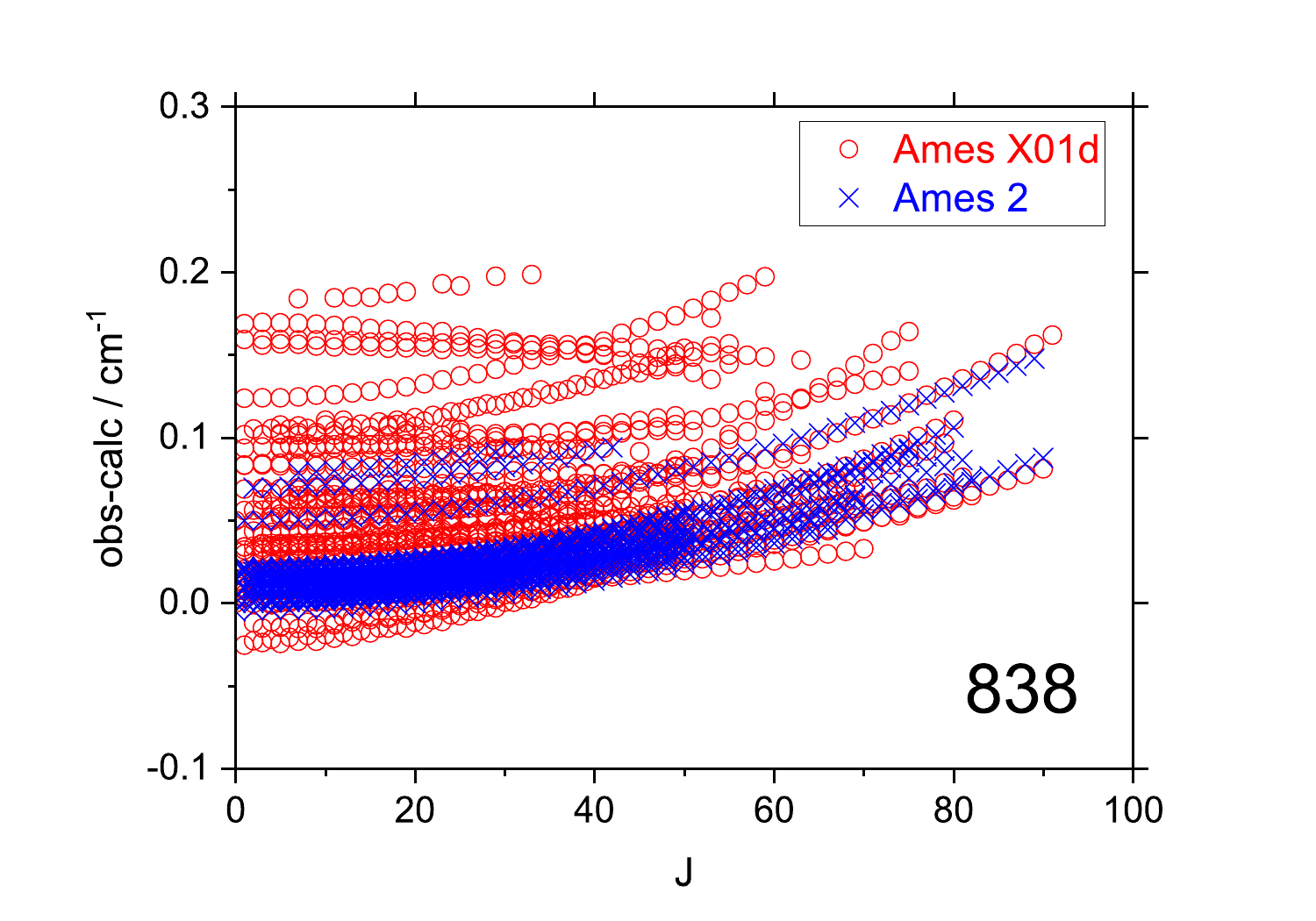}
\caption{\label{f:CO2:obs-calc:626:636} Obs.-Calc. residuals for  \COO{2}{6} and \COO{3}{8} using two Ames PESs, Ames-2 \citep{17HuScFr.CO2}, shown with red circles, and Ames-X01d \citep{23HuFrTa.CO2}, shown with blue crosses.}
\end{figure}

We initially considered using the more recent empirical PES Ames-X01d of CO$_2$ developed by \citet{23HuFrTa.CO2}, which was optimised to treat all isotopologues on an equal footing. The earlier PES, Ames-2 \citep{17HuScFr.CO2}, originally developed for the main isotopologue \COO{2}{6}, was also available and had previously been used to construct the UCL-4000 line list \citep{jt804}. However, out tests suggested that the optimisation in Ames-X01d came at the cost of somewhat degraded line list quality.
For \COO{2}{6} and \COO{3}{8}, this is illustrated in Fig.~\ref{f:CO2:obs-calc:626:636}, which compares \TROVE-calculated term values with experimentally derived values (MARVEL or CDSD-2024-PI \citep{25KoPe.CO2}) using both PESs. Ames-2 provides a more compact and flatter $J$-dependence, indicating more accurate equilibrium values and greater suitability for subsequent band-centre corrections. Our tests confirmed this trend across all 12 isotopologues, with Ames-2 consistently giving better agreement with \MARVEL\ and HITRAN energies, at least using \TROVE. We therefore adopted Ames-2 for all line lists reported in this work.

\subsection{Dipole Moment Surface}

For transition intensities we used the \textit{ab initio} DMS Ames-2021-40K of CO$_2$ by \citet{23HuFrTa.CO2}, the latest in the Ames series of dipole moment surfaces. In our previous line list, UCL-4000, the DMS of \citet{jt613}
 was employed.

\subsection{Variational nuclear-motion calculations}

The rovibrational Schr\"{o}dinger equation was solved using the program \TROVE\ \citep{TROVE_prog}. Our methodology follows \citet{20YuMexx} and \citet{jt951}, employing the exact KEO in the bisector frame together with basis functions constructed from associated Laguerre polynomials. For full details the reader is referred to these works; here we summarise only the main steps.

First, vibrational $J=0$ eigenfunctions $\Psi_{\lambda,L}^{(J=0,\Gamma_{\rm vib})}$ were obtained variationally for $L=0\ldots L_{\rm max}$ by diagonalising the $J=0$ Hamiltonian $\hat{H}^{(J=0)}$ in all relevant irreducible representations ($A_1$, $A_2$, $B_1$, $B_2$ or $A'$, $A''$ for symmetric or asymmetric species, respectively). The primitive basis was constructed as a product of 1D vibrational functions $\phi_{n_1}(r_1)$, $\phi_{n_2}(r_2)$, and $\phi_{n_3}^L(\rho)$, where $r_1$ and $r_2$ are stretching coordinates and $\rho=\pi-\alpha$ with $\alpha$ the bond angle \citep{20YuMexx}.

For $J>0$, rovibrational basis functions were formed as contracted, symmetrised products of the $J=0$ vibrational functions and rigid-rotor functions:
\begin{equation}
\label{e:Psi-basis}
\Psi_{\lambda,K}^{(J,\Gamma)} = \{ \Psi_{\lambda,K}^{(J=0,\Gamma_{\rm vib})} \, \ket{J,K,\Gamma_{\rm rot}} \}^{\Gamma},
\end{equation}
where $\ket{J,K,\Gamma_{\rm rot}}$ are symmetrised rigid-rotor functions \citep{17YuYaOv}, $K$ is constrained by $L$ ($K=L$), and $\Gamma$, $\Gamma_{\rm vib}$, $\Gamma_{\rm rot}$ are the total, vibrational, and rotational symmetries in C$_{2v}$(M) or C$_s$(M). Nuclear masses adopted were: $m_{\rm C}=11.99670909$~Da ($^{12}$C) and 13.000063355~Da ($^{13}$C); $m_{\rm O}=15.990525980$~Da ($^{16}$O), 16.99474312~Da ($^{17}$O), and 17.99477097~Da ($^{18}$O).

Rovibrational wavefunctions were computed up to $J_{\rm max}$ (see Table~\ref{t:statistics}) and used to generate transition intensities for all dipole-allowed transitions. For \COO{2}{6}, lower and upper state energies were limited to 16\,000~cm$^{-1}$ and 36\,000~cm$^{-1}$, respectively. For \COO{3}{6}, the lower-state energy cut-off was set to 12 000~cm$^{-1}$. For all other isotopologues the lower-state cut-off was 10\,000~cm$^{-1}$. Transition wavenumbers spanned 0--20\,000~cm$^{-1}$. Nuclear spin statistical weights $g_{\rm ns}$ for all isotopologues are given in Table~\ref{t:gns}.

\subsection{Empirical corrections of $J=0$ energies}

\begin{figure}
\centering

\includegraphics[width=1\textwidth]{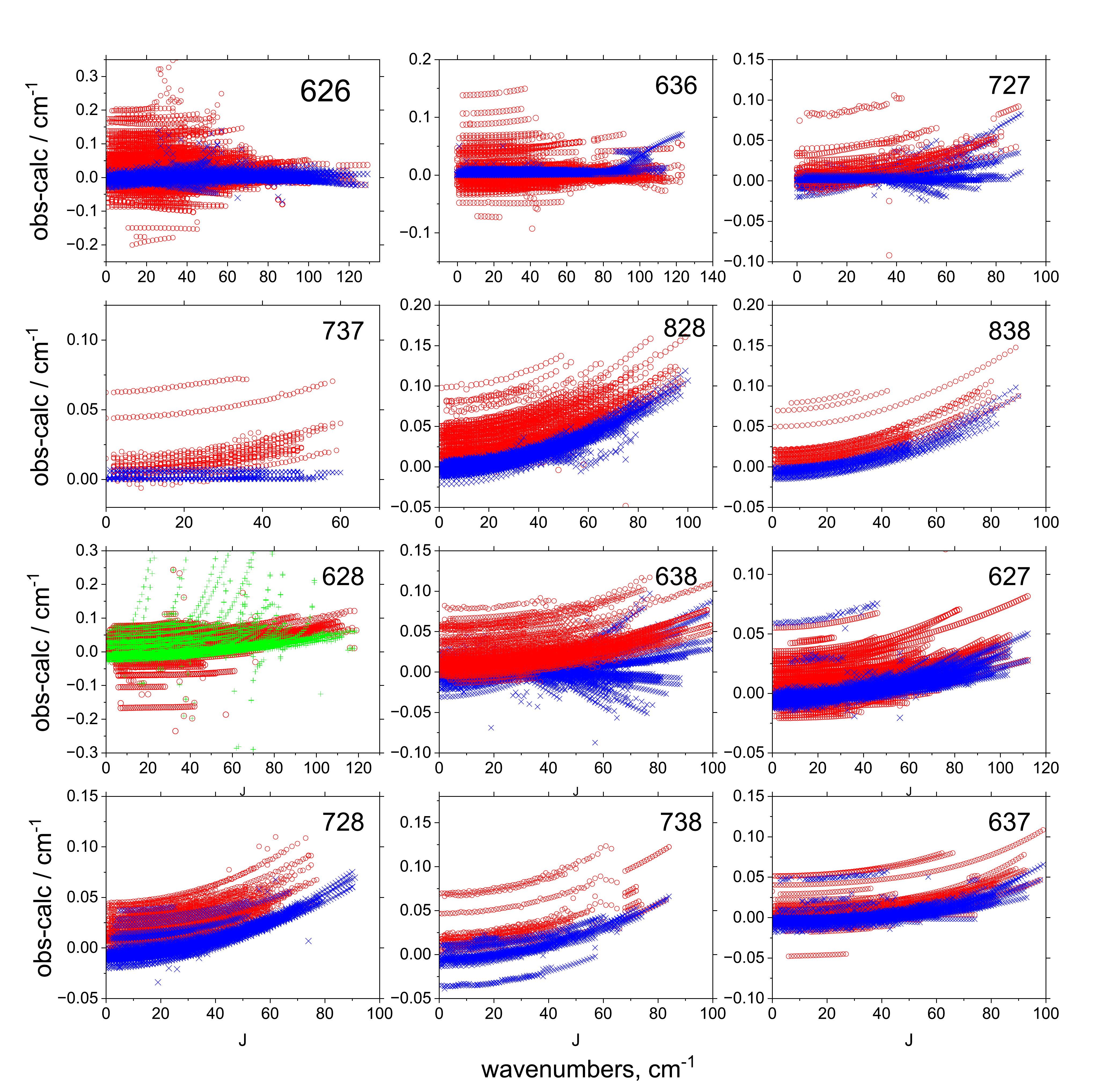}

\caption{\label{f:CO2:obs-calc:bc} `Obs.-calc'. residuals  for different isotopologues using two Ames PESs, Ames-2 \citep{17HuScFr.CO2}  and Ames-X01d \citep{23HuFrTa.CO2}, before (red circles) and after (blue crosses) the band centre corrections  \citep{09YuBaYa.NH3}. The green crosses in the display 628 show the divergence of the band-centre-corrected energies with $J$ increases (see text).
}
\end{figure}

To improve calculated rovibrational energies, empirical band-centre corrections (EBCC) \citep{09YuBaYa.NH3,jt503} were applied to the diagonal elements of the $J=0$ Hamiltonian matrices. These corrections were obtained as average residuals between \TROVE-calculated $J=0$ band origins and empirical values (from \MARVEL, CDSD-2024-PI, or HITRAN2020). For \COO{2}{6}, the procedure is illustrated in Fig.~\ref{f:CO2:obs-calc:bc}.

Since $J=0$ basis functions are eigenfunctions of the vibrational Hamiltonian, their diagonal elements can be straightforwardly replaced by empirical energies, thereby propagating improvements to higher-$J$ rovibrational levels \citep{09YuBaYa.NH3,jt729}. These also include states with $\ell_2>0$, which formally do not exist in Nature but are present in the basis set.
This significantly reduced obs.-calc. residuals across most isotopologues. The exception was \OCO{6}{2}{8}, where some bands diverged at high $J$ after EBCC (Fig.~\ref{f:CO2:obs-calc:bc}, green crosses). For this isotopologue, band-centre shifts were not applied.


\subsection{MARVELisation procedure}

Further improvements were made using the hybrid MARVELisation approach \citep{jt948}, in which calculated energies are replaced, where available, by more accurate values derived from \MARVEL\ or effective Hamiltonian (EH) analyses.
\MARVEL\ energies were taken from the recent \MARVEL\ studies of CO$_2$ isotopologues \citep{jt925,jt932,jt955,jt963,jt973,jt974,jt983,jt984}, while effective Hamiltonian energies were taken from CDSD-2024-PI  \citep{25KoPe.CO2} (\COO{2}{6}) and  HITRAN2020 \citep{jt836} (all minor isotopologues).  The EH energies are used to  minimise gaps in available \MARVEL\ energy levels.

To this end, the EH energies were generated using the \MARVEL\ algorithm applied to HITRAN transitions \citep{07FuCsTe,20ToFuSiCs,20ArFuCs.marvel,24ArFuCs}. The HITRAN uncertainness were set based on the corresponding HITRAN error code \citep{jt836}, for example, for code 4 (0.0001 \cm\ $ \leq \epsilon < $ 0.001~\cm), the uncertainty of 0.0005~\cm\ was assumed.
For the main isotopologue \COO{2}{6}, the CDSD-2024-PI energies were readily provided by \citet{25KoPe.CO2}.  The \MARVEL\ procedure ensures self-consistency of the derived network, with uncertainties propagated through the bootstrap algorithm \citep{jt908}. While \MARVEL\ is usually applied to directly measured transitions, here it is used on effective Hamiltonian-derived frequencies to supplement the empirical dataset.

In the ExoMol states files, purely \MARVEL-derived levels are labelled ``Ma'', HITRAN-derived levels as ``HI'', and effective Hamiltonian values as ``EH'' (see the description of the hybrid MARVELisation approach of \citet{jt948}). For the main isotopologue \COO{2}{6}, \MARVEL\ data from \citet{jt963} were supplemented by CDSD-2024-PI \citep{25KoPe.CO2}. The HITRAN energies generated using the \MARVEL\ approach are provided as part of the supplementary material.

Table~\ref{t:statistics} also compares the number of lines for each isotopologue with a transition intensity greater
than $10^{-30}$ cm/molecule at 296 K. For some reason which remains unclear, given that our computed line lists are
complete, HITRAN contains slightly more lines for each isotopologue.

\begin{table*}
\centering
\caption{\label{t:statistics}
Summary of CO$_2$ isotopologue line lists.
$L_{\rm max}$ is the maximum vibrational angular momentum;
$J_{\rm max}$ the maximum rotational quantum number;
$E''_{\rm max}$ the maximum lower-state energy;
$N_{\rm state}$ the number of states in the ExoMol \texttt{.states} file;
$N_{\rm lines}$ the number of transitions in the \texttt{.trans} file;
$N_{\rm HI}$ the number of HITRAN lines at $T=296$~K;
$N_{\rm Ca}$ the number of calculated ExoMol lines at $T=296$~K;
$N_{\rm Ma}$ the number of substituted lines with \MARVEL\ values at $T=296$~K;
$N_{\rm Ma+HI}$ the total number of `MARVELised' lines (Ma/HI or EH) at $T=296$~K;
and the terrestrial isotopic abundance from HITRAN \citep{jt836} is given in the final column. Only transitions stronger than  $I = 10^{-30}$ cm/molecule) at $T=$ 296~K are counted. }
\scriptsize
\begin{tabular}{rcccrrrrrrr}
\hline
Iso & $L_{\rm max}$ & $J_{\rm max}$ & $E''_{\rm max}$ & $N_{\rm state}$ & $N_{\rm lines}$ & $N_{\rm HI}$ & $N_{\rm Ca}$ & $N_{\rm Ma}$ & $N_{\rm MA+HI}$ & Abundance \\
    &               &               & (cm$^{-1}$)     &                 &                 & 296~K       & 296~K       & 296~K    & 296~K           & (HITRAN) \\
\hline
   626 &   30 &    250 &  16000 & 3\,646\,814 & 3\,235\,666\,965  &   175891 &     174445 &   175109   &     175109 & 9.8420$\times10^{-1}$  \\
   627 &   16 &    150 &  10000 & 6\,535\,175 & 1\,042\,026\,128  &    73221 &      72228 &   36582    &      73082 & 7.3399$\times10^{-4}$  \\
   628 &   20 &    150 &  10000 & 7\,369\,500 & 1\,112\,362\,652  &   116222 &     118701 &   58841    &     115774 & 3.9471$\times10^{-3}$  \\
   636 &   30 &    150 &  12000 & 2\,670\,885 &    753\,056\,397  &    70884 &      69870 &   38525    &      70779 & 1.1057$\times10^{-2}$  \\
   637 &   16 &    150 &  10000 & 7\,026\,133 & 1\,169\,286\,790  &    22830 &      22577 &    4572    &      22812 & 8.2462$\times10^{-6}$  \\
   638 &   12 &    150 &  10000 & 4\,072\,374 & 1\,246\,979\,498  &    40304 &      39950 &   12592    &      40240 & 4.4345$\times10^{-5}$  \\
   727 &   20 &    150 &  10000 & 5\,168\,133 &    583\,350\,654  &     6558 &       6493 &    5372    &       6551 & 1.3685$\times10^{-7}$  \\
   728 &   16 &    150 &  10000 & 6\,897\,362 & 1\,141\,805\,035  &     5031 &      14378 &    4186    &       5030 & 1.3685$\times10^{-7}$  \\
   737 &   20 &    150 &  10000 & 5\,133\,983 &    651\,777\,101  &     1639 &       1501 &    903     &       1533 & 1.5375$\times10^{-9}$  \\
   738 &   16 &    150 &  10000 & 7\,432\,457 & 1\,273\,579\,199  &     3580 &       3569 &    1288    &       3578 & 1.6535$\times10^{-8}$  \\
   828 &   20 &    150 &  10000 & 2\,527\,339 &    321\,570\,612  &    10541 &       1049 &    6267    &      10520 & 3.9573$\times10^{-6}$  \\
   838 &   20 &    150 &  10000 & 2\,721\,734 &    361\,867\,203  &     3053 &       2926 &    1146    &       2966 & 4.4460$\times10^{-8}$  \\
   \hline
\end{tabular}
\end{table*}

\section{Rovibrational line lists for CO$_2$ isotopologues}

Using the \TROVE\ rovibrational wavefunctions and the \textit{ab initio} DMS Ames-2021-40K, we generated 12 line lists for CO$_2$ isotopologues, collectively referred to as \name, in the ExoMol format \citep{jt939}. The coverage extends from 0 to 20\,000~cm$^{-1}$ (i.e. wavelengths $\geq 0.5~\mu$m). For the parent and most important isotopologue, \COO{2}{6}, we adopted a more extensive coverage of state excitations: a higher lower-state energy cut-off $E''_{\rm max}$ of 16\,000~cm$^{-1}$ (cf. 10\,000-12\,000~cm$^{-1}$ for other isotopologues). A larger $E''_{\rm max}$ increases computational cost but improves population coverage of excited states and thus yields a larger, more complete line list (see Table~\ref{t:statistics}). Further control of the excitation coverage is provided by $J_{\rm max}$ and $L_{\rm max}$, which were selected per isotopologue (Table~\ref{t:statistics}). The choice of $L_{\rm max}$ affects the size of the basis set and thus its quality.

Each line list comprises a \texttt{.states} file, a set of \texttt{.trans} files, and a partition-function file \texttt{.pf}. As an example, an extract from the \COO{2}{6} \texttt{.states} file is given in Table~\ref{t:states}. It contains state term values $\tilde{E}_i$, total degeneracies $g_i$, uncertainties (\texttt{unc}), lifetimes, and quantum numbers. Vibrational labels are provided in three conventions: Herzberg ($v_1$, $v_2^{\ell_2}$, $v_3$), AFGL ($m_1$, $m_2$, $l_2$, $m_3$, $r$), and \TROVE\ local-mode indices ($n_1$, $n_2$, $n_3$, $L$). Both rotationless parity ($e/f$) and molecular symmetry ($A_1$, $A_2$, $B_1$, $B_2$, $A'$, $A''$) are provided. States are indexed by an integer $i$ (state ID), used in the transition files.

The \texttt{.states} files are MARVELised: where available, calculated term values are replaced by empirical values, \MARVEL\ (\texttt{Ma}), HITRAN (\texttt{HI}) or effective-Hamiltonian (CDSD; \texttt{EH}), with the label supplied in the penultimate column; the original calculated value is retained in the final column. State-dependent lifetimes computed with \name\ are included (column 6; see Table~\ref{t:states}).

For practicality, each isotopologue's transition set is divided into 20 \texttt{.trans} files, each covering a 1000~cm$^{-1}$ bin over 0-20\,000~cm$^{-1}$. An extract from a \COO{2}{6} \texttt{.trans} file is shown in Table~\ref{t:trans}, listing Einstein A coefficients and the upper/lower state IDs.

\subsection{Using machine learning to reconstruct spectroscopic quantum numbers}

In order to help with the usage of the Dozen line list,  quantum numbers for all three assignment schemes are provided as  part of the States file. While the \TROVE\ quantum numbers are generated using the largest contribution to the associated eigenfunctions, to assign vibrational quantum numbers for AFGL and Herzberg they are not known, we developed a machine learning (ML) pipeline. To this end, a classifier was implemented in Pytorch \citep{19PaGrMa_Pytorch.ML} using a multi-head neural network architecture containing a shared backbone and specialized output heads.

The ML model starts with a shared feature extraction network consisting of 6 fully connected layers (512 $\to$ 256 $\to$ 128 $\to$ 128 $\to$ 64 neurons) with GELU activations and dropout regularization to learn common representations from 32 molecular features (energy, quantum numbers, dipole moments, isotopic masses, symmetries). From the 64-dimensional shared representation, 8 separate classification heads branch out, each containing 3 additional layers (64 $\to$ 32 $\to$ output classes) to predict the specific quantum numbers (Herzberg: $v_1$, $v_2$, $l_2$, and $v_3$ and AFGL: $m_1$, $m_2$, $m_3$, and $r$). This design enables the backbone to capture shared molecular physics, while each head specializes in its respective quantum-number assignment task. Crucially, we enforce quantum-mechanical constraints within the loss function, limiting $l_2$ values according to the parity of $v_2$.

We assess model uncertainty using Monte Carlo (MC) Dropout \citep{16GaGh_MCDrop.ML}, which keeps dropout layers active during inference by effectively sampling multiple network realizations. We perform 50 forward passes per sample, obtaining a distribution of class probabilities. From the mean probability vector $\bar{p}$, we compute predictive entropy: $H(\bar{p})=-\sum  \bar{p} \log \bar{p}$, normalized to $[0,1]$ by dividing by $\log(n_{\rm classes})$. A normalized entropy threshold of 0.4, determined from prior validation to optimize the trade-off between accepted correct predictions and rejected incorrect ones, was applied. Predictions below this threshold were accepted, enabling the assignment of over one million spectral lines across all targeted isotopologues.

\begin{figure}
\centering
\includegraphics[width=0.80\textwidth]{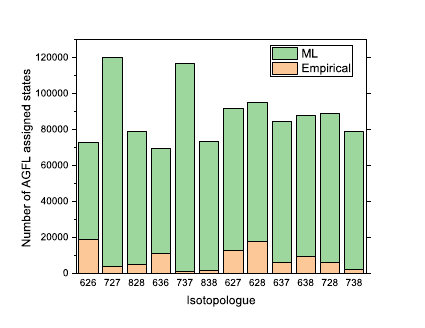}
\caption{\label{f:ML} Coverage of the assigned states with the AFGL quantum numbers per isotopologue comparing to the corresponding empirical assignments: \MARVEL, HITRAN or CDSD-2024-PI. }
\end{figure}

For different isotopologues, the ML procedure resulted in 80~000--120~000 states additionally  assignments with AFGL and Herzberg. While this represents only 1.4-3\% of the total number of states in each data set, this is a multifold increase  comparing to the empirically assignment states, which is illustrated in Fig.~\ref{f:ML}.

\subsection{Partition functions}

\begin{figure}
\centering
\includegraphics[width=0.80\textwidth]{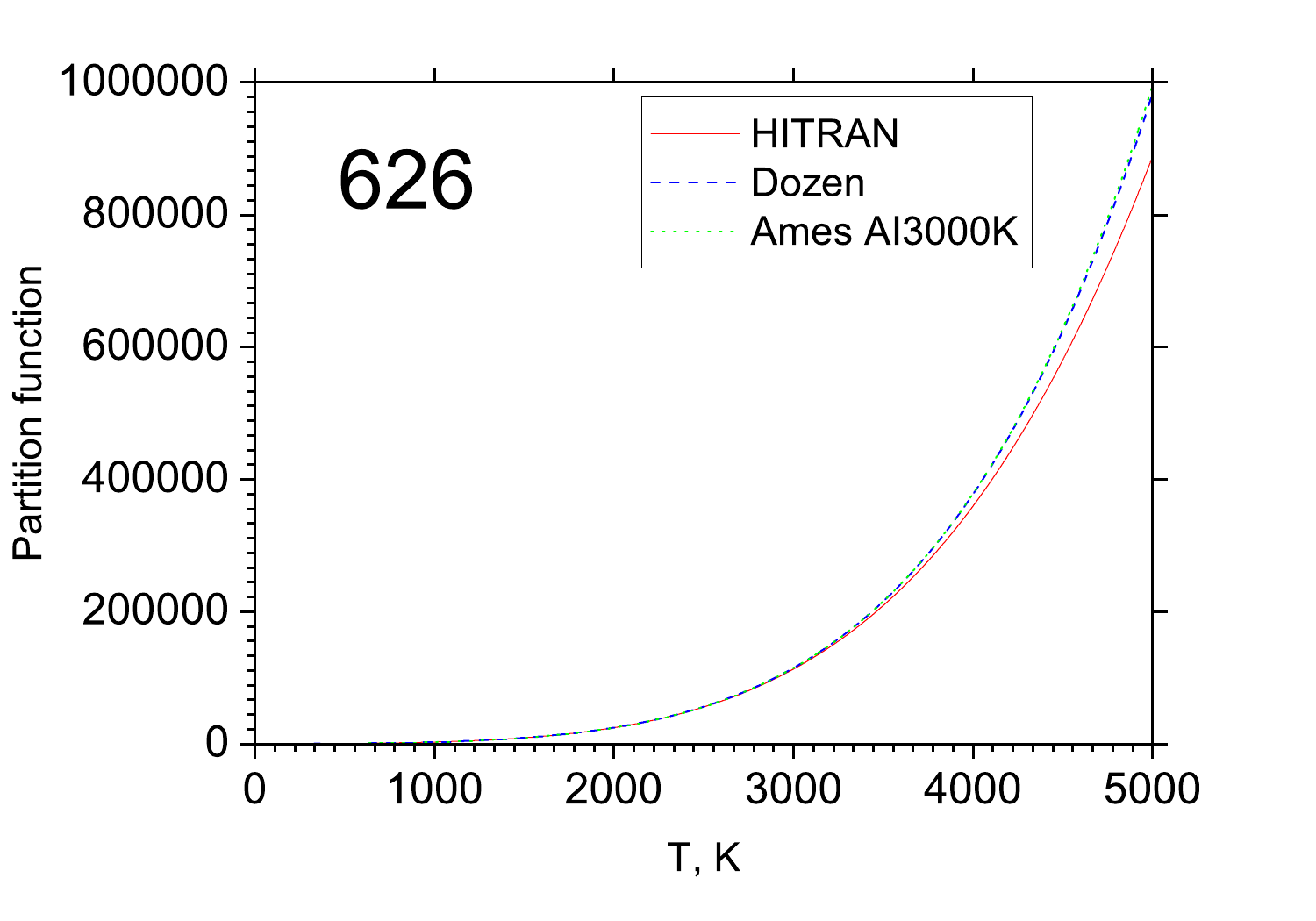}
\caption{\label{f:pf:626} Partition functions for \COO{2}{6}: comparison of \name\ with TIPS 2024  of HITRAN2024 \citep{jt981} and Ames AI-3000K.}
\end{figure}

Partition functions were computed for all 12 isotopologues on a 1~K grid from 0 to 5000~K using the direct sum
\[
Q(T) = \sum_{i} g_{\rm ns}^{(i)}\, \bigl(2J_i+1\bigr)\, \exp\!\left(-\frac{c_2\,\tilde{E}_i}{T}\right),
\]
where $g_{\rm ns}^{(i)}$ is the nuclear-spin statistical factor (Table~\ref{t:gns}), $J_i$ is the rotational quantum number, $c_2$ is the second radiation constant (cm\,K), and $\tilde{E}_i$ is the energy term value (cm$^{-1}$)  relative to the ground state ($J=0$, $v_1=v_2=v_3=0$, $\ell_2=0$, $e$).

Figure~\ref{f:pf:626} compares the \name\ partition function for \COO{2}{6} with TIPS 2024 (total internal partition sum) of HITRAN2020 \citep{jt981} and with that obtained from the Ames AI-3000K line list \citep{23HuFrTa.CO2}. The \name\ and Ames curves agree closely and both lie systematically above the HITRAN TIPS values, reflecting improved completeness of the underlying states; the lack of convergence of the HITRAN TIPS partition function at high temperature has been noted previously \citep{jt899}.

\begin{table*}
\centering
\caption{\label{t:states} Extract from the \texttt{.states} file of \COO{2}{6}  \name\ line list. \ }
{\setlength{\tabcolsep}{1.5pt}
\tt
\begin{tabular}{rrrrrrcrrrrrcrrrrrrrrrrrrrrrrrrrrr}
\toprule \toprule
        $i$  &  \multicolumn{1}{c}{$\tilde{E}$/\cm}   &  $g$  &  $J$  & \multicolumn{1}{c}{unc/\cm} & \multicolumn{1}{c}{$\tau$/s} & $\Gamma_{\rm tot}$ &$e/f$& $v_1$ & $v_2$ & $l_2$ & $v_3$ & $|C_i^{2}|$ &$m_1$ & $m_2$ & $l_2$ & $m_3$ & $r$ & $n_1$ & $n_2$ & $n_3$ & \multicolumn{1}{c}{Label}  &$\tilde{E}_{\rm calc.}$/\cm  \\
 \midrule
       1 &      0.000000 &   1 &    0 &  0.000001 & inf         & A1  & e   &   0 &   0 &   0 &   0 & 1.00 &   0 &  0 &  0 &   0 &   1 &   0 &   0 &   0 & Ma  &       0.000000\\
       2 &   1285.408294 &   1 &    0 &  0.000049 &  7.7985E-01 & A1  & e   &   0 &   2 &   0 &   0 & 0.54 &   1 &  0 &  0 &   0 &   2 &   0 &   0 &   1 & Ma  &    1285.407001\\
       3 &   1388.184102 &   1 &    0 &  0.000053 &  5.5304E-01 & A1  & e   &   1 &   0 &   0 &   0 & 0.54 &   1 &  0 &  0 &   0 &   1 &   1 &   0 &   0 & Ma  &    1388.184019\\
       4 &   2548.363900 &   1 &    0 &  0.001515 &  3.5601E-01 & A1  & e   &   1 &   2 &   0 &   0 & 0.47 &   2 &  0 &  0 &   0 &   3 &   1 &   0 &   1 & Ma  &    2548.364344\\
       5 &   2671.143088 &   1 &    0 &  0.001119 &  4.8854E-01 & A1  & e   &   2 &   0 &   0 &   0 & 0.66 &   2 &  0 &  0 &   0 &   2 &   1 &   1 &   0 & Ma  &    2671.142828\\
       6 &   2797.136399 &   1 &    0 &  0.001000 &  2.3234E-01 & A1  & e   &   1 &   2 &   0 &   0 & 0.52 &   2 &  0 &  0 &   0 &   2 &   1 &   0 &   1 & EH  &    2797.135151\\
       7 &   3792.682402 &   1 &    0 &  0.001000 &  2.3664E-01 & A1  & e   &   1 &   4 &   0 &   0 & 0.43 &   3 &  0 &  0 &   0 &   3 &   1 &   0 &   2 & EH  &    3792.680795\\
       \multicolumn{3}{c}{\ldots} & \multicolumn{3}{c}{\ldots} & \multicolumn{3}{c}{\ldots} & \multicolumn{3}{c}{\ldots} \\
      18 &   6016.690108 &   1 &    0 &  0.001000 &  1.2107E-03 & A1  & e   &   1 &   0 &   0 &   2 & 0.60 &   1 &  0 &  0 &   2 &   0 &   0 &   3 &   0 & EH  &    6016.701190\\
      19 &   6240.057942 &   1 &    0 &  0.080000 &  1.5052E-01 & A1  & e   &   2 &   6 &   0 &   0 & 0.30 &   5 &  0 &  0 &   0 &   4 &   1 &   1 &   3 & Ca  &    6240.057942\\
      20 &   6435.500064 &   1 &    0 &  0.060000 &  1.7500E-01 & A1  & e   &   4 &   2 &   0 &   0 & 0.28 &   5 &  0 &  0 &   0 &   4 &   2 &   2 &   1 & Ca  &    6435.500064\\
      21 &   6588.319980 &   1 &    0 &  0.050000 &  2.2609E-01 & A1  & e   &   5 &   0 &   0 &   0 & 0.48 &   5 &  0 &  0 &   0 &   2 &   3 &   2 &   0 & Ca  &    6588.319980\\
       \multicolumn{3}{c}{\ldots} & \multicolumn{3}{c}{\ldots} & \multicolumn{3}{c}{\ldots} & \multicolumn{3}{c}{\ldots} \\
      30 &   7834.845932 &   1 &    0 &  0.040000 &  1.7933E-01 & A1  & e   &   1 &   0 &   0 &   3 & 0.29 &  -1 & -1 & -1 &  -1 &  -1 &   2 &   2 &   0 & Ca  &    7834.845932\\
      31 &   7974.351788 &   1 &    0 &  0.040000 &  1.7354E-01 & A1  & e   &   1 &   0 &   0 &   3 & 0.51 &  -1 & -1 & -1 &  -1 &  -1 &   2 &   2 &   0 & Ca  &    7974.351788\\
\toprule \toprule
\end{tabular}}
\mbox{}\\
{\flushleft
\begin{tabular}{ll}
\toprule
$i$:& State identifier (integer index used in \texttt{.trans} files). \\
$\tilde{E}$:& State term value (cm$^{-1}$). \\
$g$:& Total state degeneracy. \\
$J$:& Total rotational quantum number. \\
unc:& Energy uncertainty (cm$^{-1}$). \\
$\tau$:& State lifetime (s). \\
$\Gamma_{\rm tot}$:& Total symmetry in C$_{2v}$(M) or C$_s$(M). \\
$e/f$:& Rotationless parity label. \\
$v_1,\,v_2^{\ell_2},\,v_3$:& Normal-mode vibrational quantum numbers (Herzberg notation). \\
$|C_i|^{2}$:& Largest expansion-coefficient weight used for assignment. \\
$m_1,\,m_2,\,l_2,\,m_3,\,r$:& AFGL vibrational quantum numbers ($-1$ stands for non-available). \\
$n_1,\,n_2,\,n_3,\,L$:& \TROVE\ local-mode quantum numbers. \\
Label:&  \Marvel\ (\texttt{Ma}), Effective Hamiltonian (\texttt{EH}), \\
& HITRAN (\texttt{HI}), or calculated \name\ value (\texttt{Ca}). \\
$\tilde{E}_{\rm calc.}$:& Original calculated \name\ term value (cm$^{-1}$). \\
\bottomrule
\end{tabular}
}

\end{table*}

\begin{table}
\centering
\caption{\label{t:trans}
Extract from a \texttt{.trans} file of the \COO{2}{6} \name\ line list.}
\tt
\centering
\begin{tabular}{rrrr}
\toprule\toprule
\multicolumn{1}{c}{$f$}	&	\multicolumn{1}{c}{$i$}	& \multicolumn{1}{c}{$A_{fi}$} \\
\midrule
      872090  &     849449 & 3.7641e-16 \\
     1729336  &    1740374 & 1.9742e-09 \\
      443590  &     423498 & 2.7506e-15 \\
       89228  &      95628 & 1.9493e-12 \\
     1562467  &    1577182 & 1.1399e-08 \\
     1031039  &    1081028 & 3.1880e-14 \\
     1710529  &    1691051 & 4.1274e-14 \\
     2476483  &    2488556 & 1.5718e-09 \\
     1139554  &    1151915 & 5.3921e-11 \\
     1726661  &    1675412 & 4.8961e-11 \\
 \bottomrule\bottomrule
\end{tabular} \\ \vspace{2mm}
\rm
\noindent
$f$: Upper  state counting number;\\
$i$:  Lower  state counting number; \\
$A_{fi}$:  Einstein A coefficient (in s$^{-1}$).
\end{table}

\subsection{Opacities}

Temperature- and pressure-dependent opacities for all 12 isotopologues of CO$_2$, based on the \name\ line lists, were generated using the ExoMolOP procedure \citep{jt801}. Opacity tables were produced for four widely used atmospheric retrieval codes: \textsc{ARCiS} \citep{ARCiS}, \textsc{TauREx} \citep{TauRex3}, \textsc{NEMESIS} \citep{NEMESIS}, and \textsc{petitRADTRANS} \citep{19MoWaBo.petitRADTRANS}, and are provided alongside the line lists. In addition, we supply combined opacities constructed for a representative atmosphere containing multiple isotopologues in terrestrial abundance ratios (see Table~\ref{t:statistics}).

\subsection{Line broadening coefficients}

Accurate treatment of collisional line broadening is essential for radiative-transfer modelling. We adopt a semi-empirical approach, combining theoretical predictions with carefully selected experimental data, to provide broadening and shifting coefficients for CO$_2$ transitions with common perturbers (N$_2$, O$_2$, CO$_2$, H$_2$, He, and H$_2$O) over the temperature range 200-2000~K.

Previous analyses have highlighted inconsistencies among experimental datasets. For example, \citet{13GaLaxx.broad} and \citet{14GaLaBl.broad} showed that different experiments on the same transition often disagree beyond the quoted uncertainties, complicating attempts to assess vibrational or isotopologue dependences of the half-width coefficients $\gamma$. More recent work by \citet{20HaGoTr.broad} re-analysed the most reliable data, proposing a new set of semi-empirical parameters for CO$_2$- and air-broadening, suitable for Voigt and speed-dependent Voigt profiles, and found no significant vibrational dependence of $\gamma_0$.

Isotopologue effects are also small. For example, \citet{98DeBeSm.broad} found no statistically significant difference between N$_2$-broadening of \OCO{6}{3}{8} and that of the main isotopologue, while \citet{13GaLaxx.broad} estimated that isotopic mass effects reduce $\gamma$ by less than 1\%. This was recently confirmed by cavity ringdown measurements of \COO{3}{6} by \citet{25MoCaGa.broad}, who found air-broadening coefficients only 0.4\% smaller than those of \COO{2}{6}.

Given the very weak vibrational and isotopologue dependences, we provide only $J$-dependent broadening coefficients $\gamma_0$ $(296~\mathrm{K})$ and temperature exponents $n$ for the single power-law parametrisation. Speed dependence and line-mixing are not considered. The adopted $\gamma_0$ and $n$ values are taken from HITRAN for broadening by H$_2$, He \citep{22TaSkSa.broad}, N$_2$, CO$_2$ \citep{jt836}, and H$_2$O \citep{19TaKoRo.broad}, and are supplied in the \textsf{m0}-diet format for use with \textsc{ExoCross} \citep{jt939}. In addition, we provide more recent theoretical water-broadening data \citep{24ViGa.broad} in the \textsf{m2}-diet format \citep{25SoYuTe}, which employs a double power-law to describe the temperature dependence of $\gamma(T)$ more accurately.

\section{Illustrations and analysis}

\begin{figure}
\centering
\includegraphics[width=0.9\textwidth]{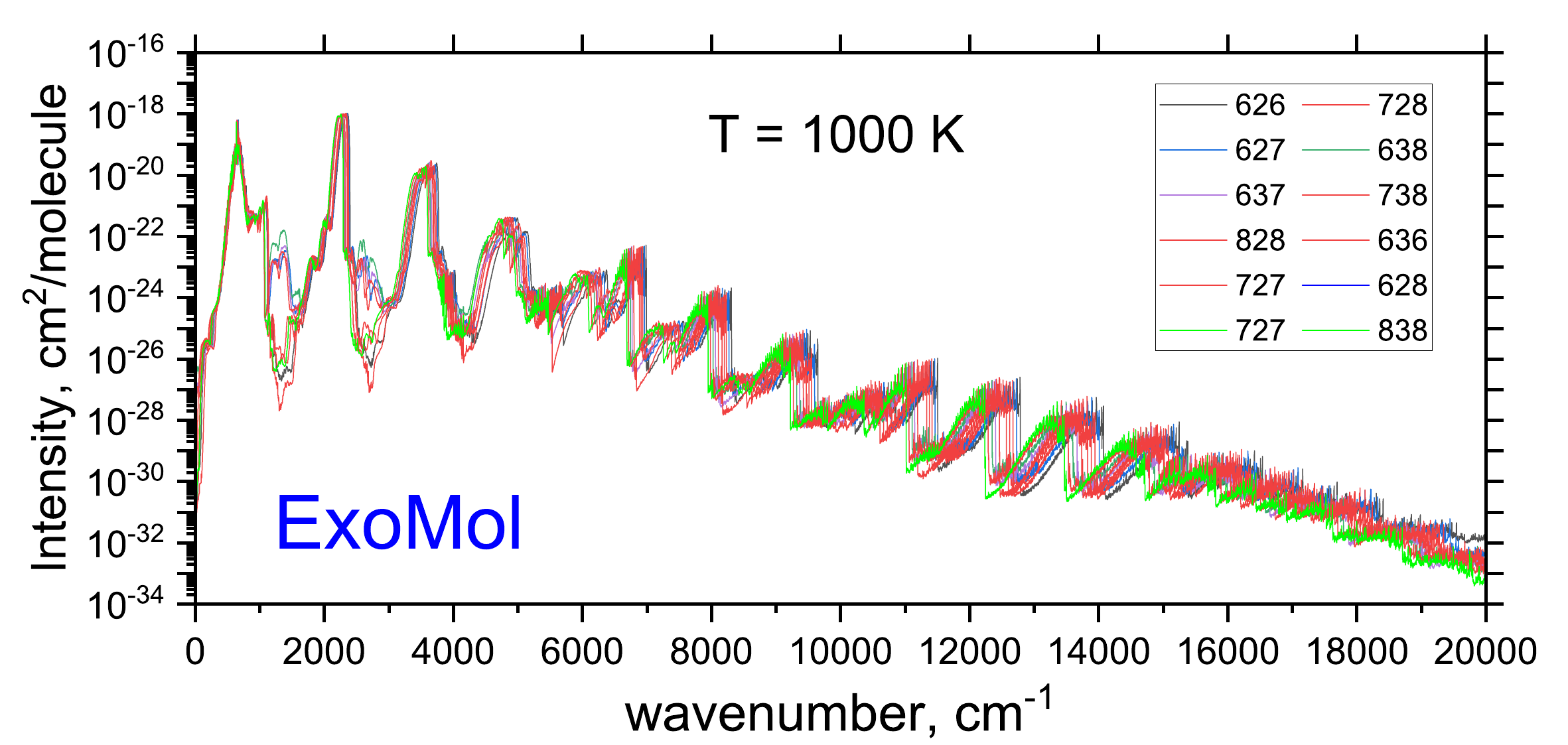}
\caption{\label{f:CO2:iso:1000K} Spectra of CO$_2$ isotopologues at $T=1000$~K (Gaussian profile, HWHM = 1~cm$^{-1}$). Intensities are not scaled to isotopic abundances.}
\end{figure}

In this section we demonstrate the performance of the new line lists by comparing their predicted spectra with previous datasets and with available experimental information. A general overview is provided in Fig.~\ref{f:CO2:iso:1000K}, which shows spectra of all isotopologues at $T=1000$~K on a logarithmic scale, without scaling to isotopic abundances.

\begin{figure}
\centering
\includegraphics[width=0.9\textwidth]{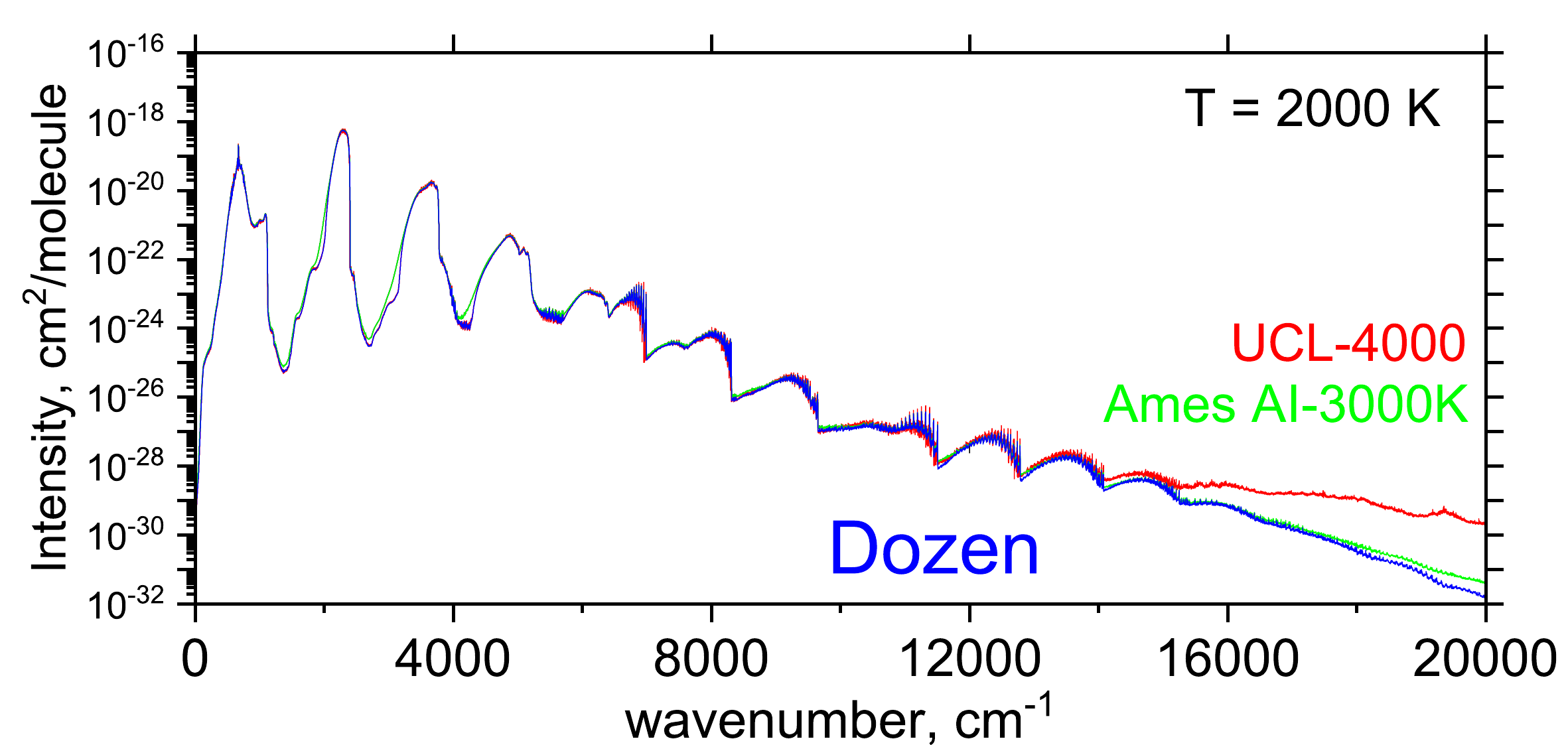}
\caption{\label{f:CO2:2020-2024} Cross-sections of \COO{2}{6} at $T=2000$~K computed with UCL-4000, \name, AI-3000K, and HITEMP~2025 (Gaussian profile, HWHM = 1~cm$^{-1}$). The HITEMP~2025  and AI-3000K \citep{23HuFrTa.CO2} spectra coincide over 0-12\,000~cm$^{-1}$.}
\end{figure}

Figure~\ref{f:CO2:2020-2024} illustrates the coverage and quality of the \name\ \COO{2}{6} line list at $T=2000$~K, compared with our previous UCL-4000 list and the Ames AI-3000K list \citep{23HuFrTa.CO2}. The \name\ results are in excellent agreement with AI-3000K. We also compared with HITEMP~2025 \citep{25HaGoHu.CO2}, which was generated from AI-3000K using the super-lines and super-energies technique \citep{20HaGoRe.CH4}. As expected, the HITEMP spectrum follows AI-3000K closely over 0-12\,000~cm$^{-1}$, which is the range adopted by HITEMP for \COO{2}{6}.

Compared to UCL-4000, the new line list exhibits more physical  behaviour at higher wavenumbers. This was achieved by optimising the stretching basis set comparing to that used in \citet{jt804} by extending the
bond-length grid used in the numerical integrations from 2.2~\AA\ to $\sim$2.6~\AA. The wavenumber behaviour of \name\ is now
consistent with the near-integrated dipole limit (NIDL) \citep{22MeUs.CO,jt794}, which predicts that overtone intensities decrease approximately exponentially, appearing as straight lines in a log-intensity plot \citep{16MeMeSt}. By contrast, the formation of plateaux at high overtones is a well-known indicator of numerical artefacts, as seen in UCL-4000 above 14\,000~cm$^{-1}$. 

\begin{figure}
\centering
\includegraphics[width=0.7\textwidth]{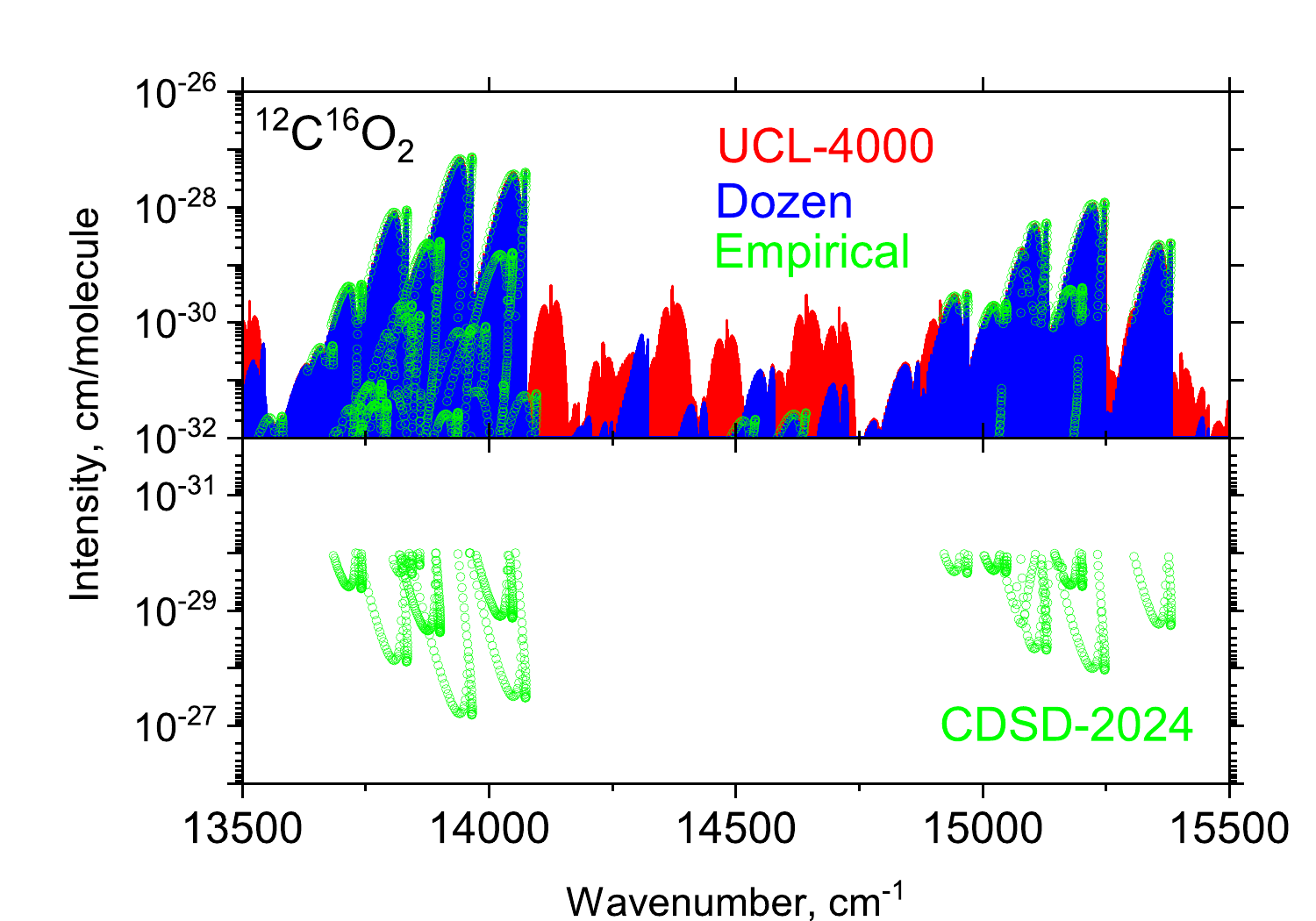}
\caption{\label{f:CO2:700nm} Intensity behaviour near 700~nm ($\sim$14\,300~cm$^{-1}$). UCL-4000 overestimates this band, while the new \name\ list based on Ames-2021-40K yields significantly lower intensities, in better agreement with the experiment of \citet{24BaRaWo.CO2}.}
\end{figure}

A more subtle intensity problem in UCL-4000 was identified by \citet{24BaRaWo.CO2}. Transitions in the 700~nm region ($\sim$14\,300~cm$^{-1}$) were predicted with intensities above $10^{-30}$~cm/molecule, yet an extensive experimental search found no evidence for them. This band was therefore significantly overestimated in UCL-4000. The present line list, based on the improved basis set and different Ames-2021-40K DMS, brings these intensities below the 10$^{-30}$ cm molecule$^{-1}$ threshold value, as shown in Fig.~\ref{f:CO2:700nm}.

\begin{figure}
\centering
\includegraphics[width=0.95\textwidth]{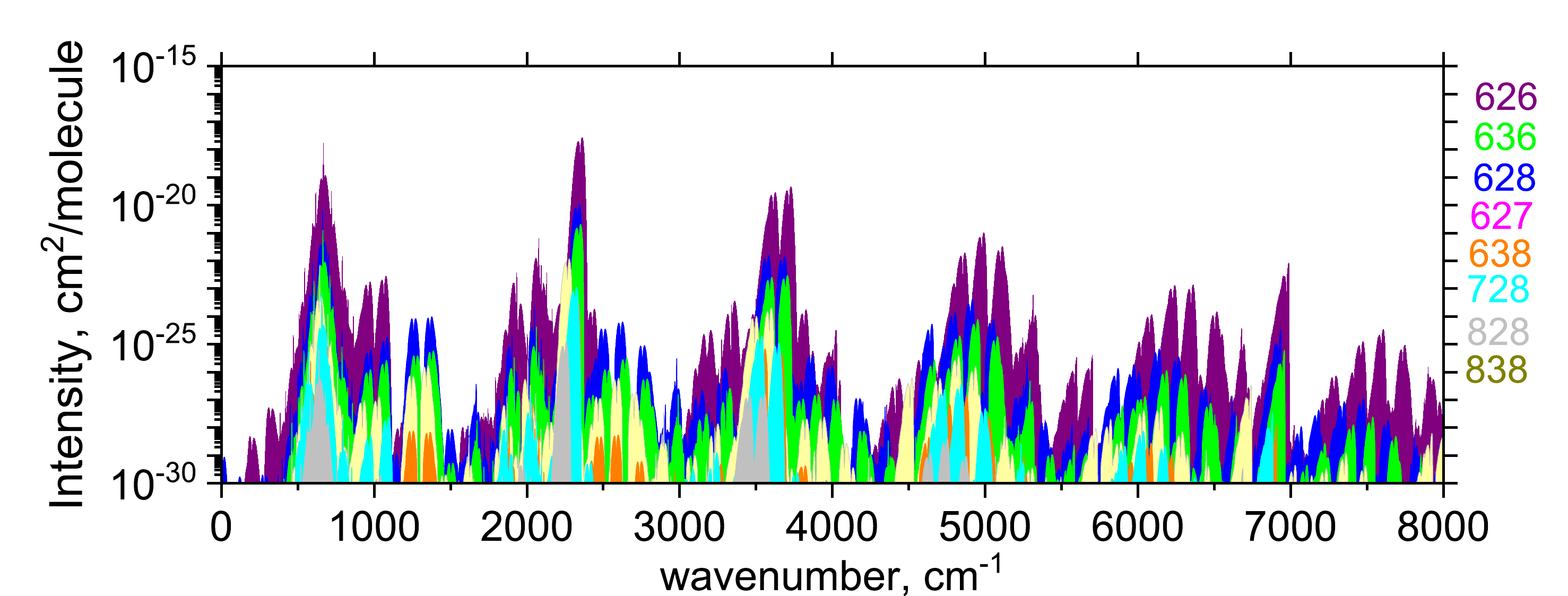}
\includegraphics[width=0.48\textwidth]{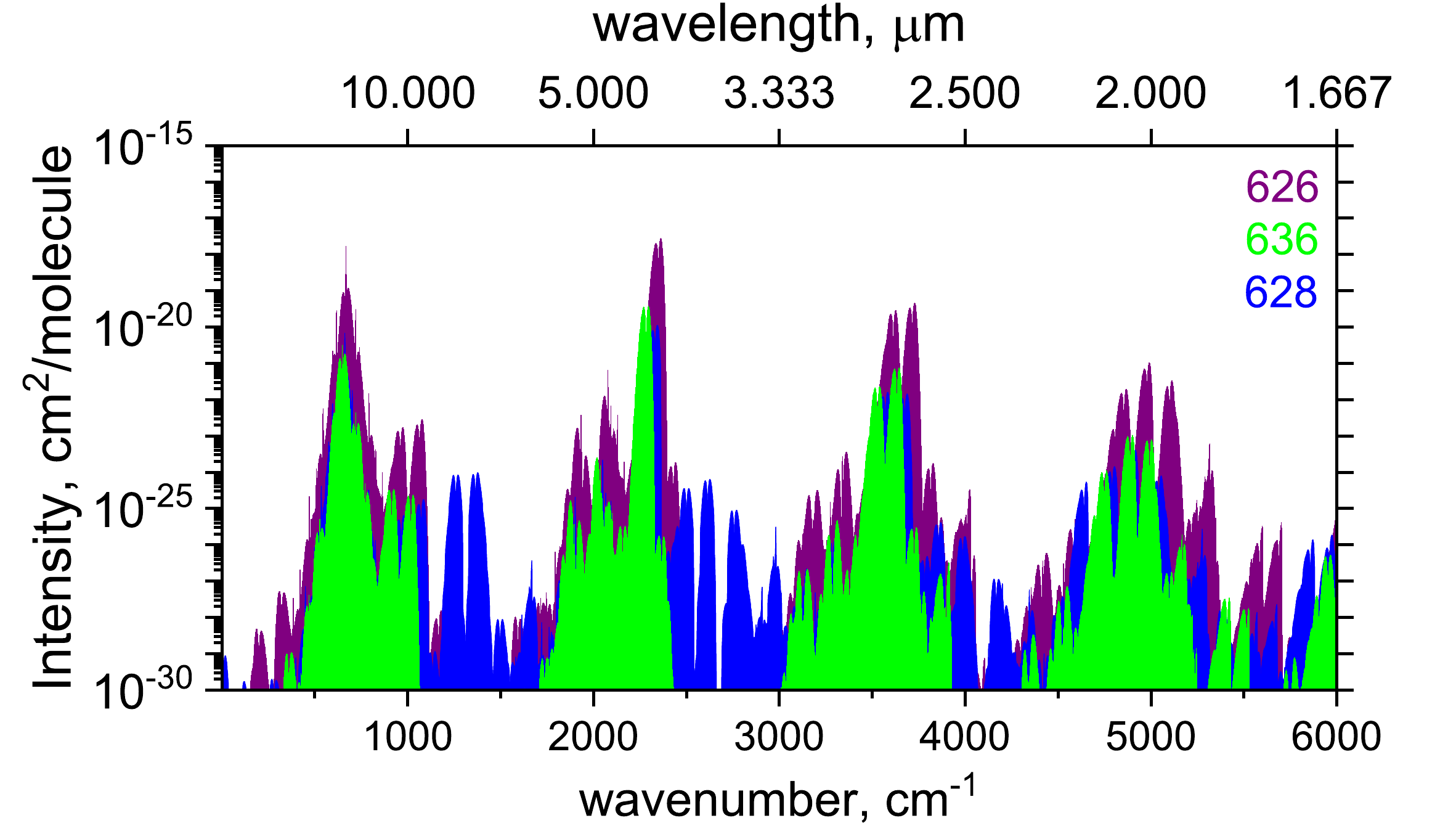}
\includegraphics[width=0.48\textwidth]{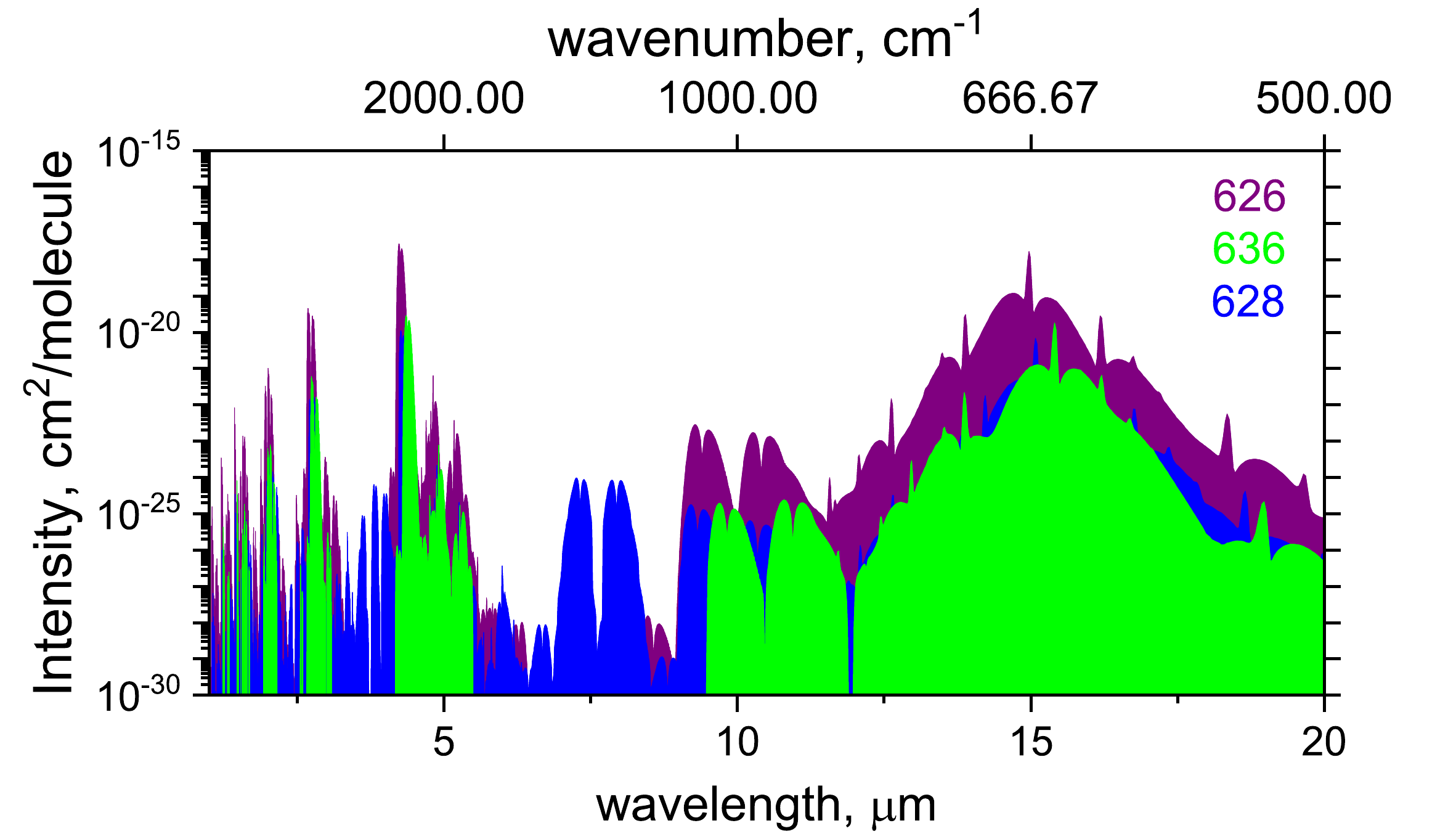}
\caption{\label{f:iso:3} Spectra of selected CO$_2$ isotopologues at $T=296$~K, scaled by terrestrial isotopic abundances. In addition to \COO{2}{6}, only \COO{3}{6} and \OCO{6}{2}{8} produce detectable IR features.}
\end{figure}

Differences among isotopologues are illustrated in Fig.~\ref{f:iso:3}, which shows $T=296$~K spectra scaled by natural terrestrial abundances. Apart from the dominant \COO{2}{6}, only \COO{3}{6} and \OCO{6}{2}{8} produce IR features of sufficient strength to be detectable under atmospheric conditions. Of course,
this situation will change with the altered isotope abundances encountered elsewhere in space.

\begin{figure}
\centering
\includegraphics[width=0.95\textwidth]{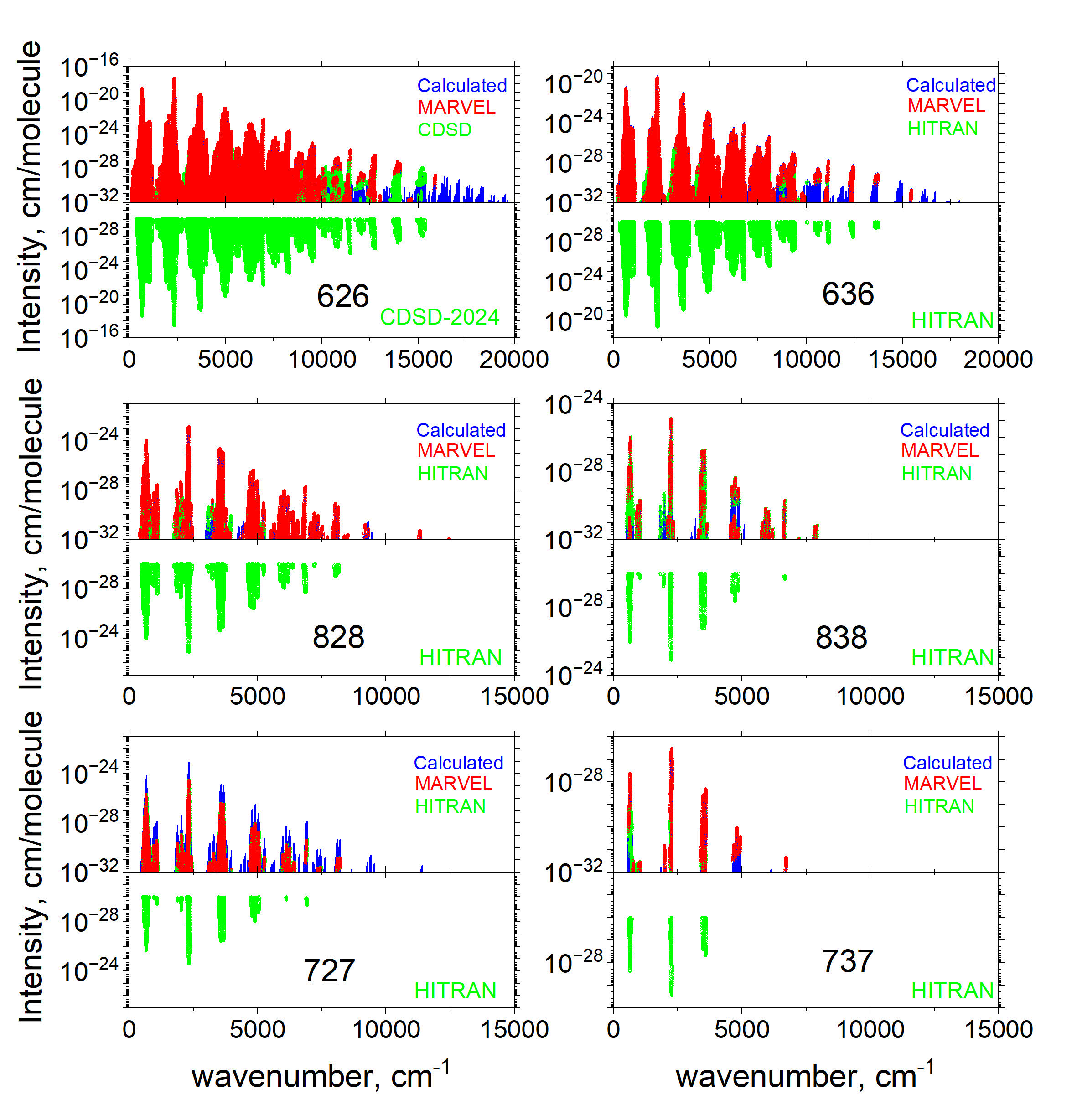}
\caption{\label{f:CO2:iso:sym:296K} Stick spectra of six symmetric isotopologues of CO$_2$ at $T=296$~K (absorption coefficients in cm/molecule). MARVELised transitions are highlighted, and comparisons are made with CDSD-2024 (\COO{2}{6}) and HITRAN2020. Intensities are not scaled to isotopic abundances.}
\end{figure}

\begin{figure}
\centering
\includegraphics[width=0.95\textwidth]{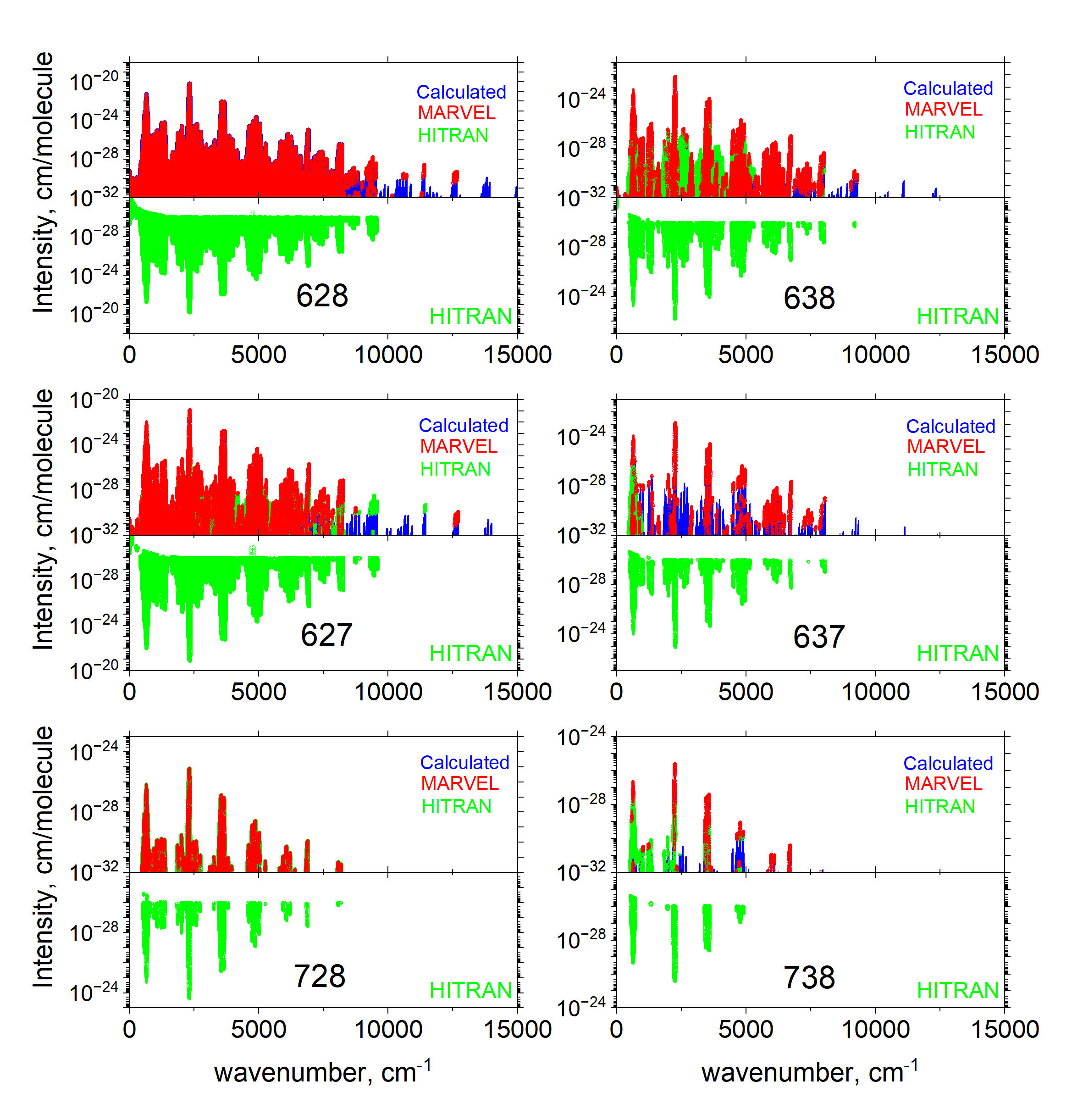}
\caption{\label{f:CO2:iso:asym:296K} Stick spectra of six asymmetric isotopologues of CO$_2$ at $T=296$~K (absorption coefficients in cm/molecule). MARVELised transitions are highlighted and compared with HITRAN2020. Intensities are not scaled to isotopic abundances.}
\end{figure}

Finally, Figures~\ref{f:CO2:iso:sym:296K} and \ref{f:CO2:iso:asym:296K} present stick spectra at $T=296$~K for all 12 isotopologues of CO$_2$, separated into symmetric and asymmetric species. MARVELised transitions are highlighted, illustrating the extent of experimental coverage. Comparisons with HITRAN2020 (and CDSD-2024 for \COO{2}{6}) further demonstrate the accuracy and completeness of the new line lists. A quantitative summary of line list statistics, including the number of empirically anchored transitions above the HITRAN intensity threshold of $10^{-30}$~cm/molecule at $T=296$~K, is given in Table~\ref{t:statistics}.

\section{Conclusions}

This work  presents a new set of comprehensive rovibrational line lists for 12 isotopologues of CO$_2$ collectively know as the \name\ line list. These line lists were computed using accurate empirical PESs and the latest \textit{ab initio} DMSs from the Ames group \citep{17HuScFr.CO2,23HuFrTa.CO2}, covering the spectroscopic range up to 20\,000~cm$^{-1}$. The line list for the main isotopologue \COO{2}{6} is applicable up to at least 3000~K, while those for the minor isotopologues are reliable up to about 2000~K. The accuracy of the datasets has been enhanced through empirical band-centre corrections and systematic MARVELisation, incorporating experimental energies from \MARVEL, HITRAN, and CDSD where available.

The new ExoMol CO\2\ line lists represent a significant improvement over previous work. In particular, we now provide hot line lists for 12 isotopologues with an extended wavenumber coverage. Besides, spectra generated using \name\
display physically correct intensity behaviour for the high overtone bands thus avoiding the spurious plateau observed in UCL-4000 \citep{jt804}. The intensity anomaly in the 700~nm region is resolved by improving the stretching basis set used. Comparisons with HITRAN spectra at $T=296$~K show excellent agreement, confirming both the reliability of line positions and the accuracy of intensities.

For the first time, we employed machine-learning methods to reconstruct AFGL and `Herzberg' quantum numbers for CO$_2$ isotopologues, significantly extending the quantum assignment coverage in a consistent and physically constrained manner. We are currently using machine-learning techniques to improve the prediction transition wavenumbers for isotopologues \citep{jtMLiso}; these results will be used to update Dozen states files for minor isotopologues with improved estimates for their energy levels in the near future. Partition functions have been computed on a fine temperature grid up to 5000~K, ensuring robust thermodynamic and opacity applications. Pressure-broadening coefficients with the most important perturbers (N$_2$, O$_2$, CO$_2$, H$_2$, He, H$_2$O, and air) have been collected from the literature and are provided in ExoMol's diet formats (\textsf{m0} and \textsf{m2}), enabling flexible use in radiative-transfer models.

Opacities were generated with ExoMolOP \citep{jt801} for four leading atmospheric retrieval codes (\textsc{ARCiS}, \textsc{TauREx}, \textsc{NEMESIS}, and \textsc{petitRADTRANS}). Combined opacity grids, scaled to terrestrial isotopic abundances, are also supplied for practical applications in planetary and stellar atmosphere modelling.

The line lists and associated data (partition functions, broadening parameters, opacities) are available in the ExoMol database at \url{www.exomol.com}. Together, they provide the most complete and accurate spectroscopic resource to date for CO$_2$ and its isotopologues. These datasets will be invaluable for atmospheric retrievals across the Solar System, brown dwarfs, and exoplanets, as well as for laboratory and terrestrial remote-sensing studies. Future work will focus on extending the spectral range into the ultraviolet and further refining high-temperature coverage.

\section*{Declaration of competing interest}
The authors declare that they have no known competing financial interests or personal relationships that could have appeared to
influence the work reported in this paper.

\section*{Acknowledgements}

This work was supported by the STFC Projects   ST/Y001508/1 and UKRI/ST/B001183/1. The authors acknowledge the use of the Cambridge Service for Data Driven Discovery (CSD3) and  the DiRAC Data Intensive service DIaL2.5 at the University of Leicester, managed on behalf of the STFC DiRAC HPC Facility (www.dirac.ac.uk). These DiRAC services were funded by BEIS, UKRI and STFC capital funding and STFC operations grants. DiRAC is part of the UKRI Digital Research Infrastructure. We thank Xinchuan Huang for help accessing AI-3000K and providing valuable suggestions. This work was also supported by the European Research Council (ERC) under the European Union's Horizon 2020 research and innovation programme through Advance Grant number 883830.

\section*{Data Availability}

All data generated in this article are available in the article and its supplementary materials
or via the the ExoMol website \href{https://www.exomol.com}{www.exomol.com}.
The HITRAN energies of 11 minor isotpologues generated using the \MARVEL\ procedure  underlying this article are available in the online supplementary material. The line lists and associated
data are available from \href{www.exomol.com}{www.exomol.com}. The codes used in this work, namely \textsc{TROVE} and \textsc{ExoCross}, are freely available via \href{https://github.com/exomol}{https://github.com/exomol}.


\bsp	
\label{lastpage}

\end{document}